\documentclass{IEEEtaes}

\usepackage{color,array,amsthm}
\usepackage{graphicx}
\usepackage{amsmath,amssymb,amsfonts}
\usepackage{algorithm}
\usepackage{array}
\usepackage{placeins}
\usepackage{algorithmic}
\usepackage{graphicx}
\usepackage{textcomp}
\usepackage{xcolor}
\usepackage{makecell}
\usepackage{comment}
\usepackage{flushend}
\usepackage{booktabs} 
\usepackage{pifont}
\usepackage{lipsum}
\usepackage{stfloats}
\usepackage[normalem]{ulem}

\jvol{XX}
\jnum{XX}
\jmonth{XXXXX}
\paper{1234567}
\pubyear{2025}
\doiinfo{TAES.2022.Doi Number}

\newcommand{\cmark}{\ding{51}}%
\newcommand{\xmark}{\ding{55}}%

\begin{document}

\title{On the Limits of Coherent Time-Domain Cancellation of Radio Frequency Interference}

\author{XINRUI LI}

\author{R. MICHAEL BUEHRER}
\member{Fellow, IEEE}
\affil{Virginia Tech, Blacksburg, VA 24061, USA}


\receiveddate{Manuscript received XXXXX 00, 0000; revised XXXXX 00, 0000; accepted XXXXX 00, 0000.\\
This work was supported in part by the National Science Foundation under Grant ECCS-2029948.}

\authoraddress{Authors’ addresses: Xinrui Li and R. Michael Buehrer are with the Wireless@VT, Bradley Department of ECE, Virginia Tech, Blacksburg, VA 24061 USA, E-mail: (lxinrui3@vt.edu; rbuehrer@vt.edu).} 
\corresp{{\itshape (Corresponding author: Xinrui Li)}. }

\markboth{LI ET AL.}{ON THE LIMITS OF CTC OF RFI}
\maketitle

\begin{abstract}
In many sensing (viz., radio astronomy) and radar applications, the received signal of interest (SOI) exhibits a significantly wider bandwidth or weaker power than the interference signal, rendering it indistinguishable from the background noise. Such scenarios arise frequently in applications such as passive radar, cognitive radio, low-probability-of-intercept (LPI) radar, and planetary radar for radio astronomy, where canceling the radio frequency interference (RFI) is critical for uncovering the SOI. In this work, we examine the Demodulation-Remodulation (Demod-Remod) based interference cancellation framework for the RFI \cite{6395266}. This approach demodulates the unknown interference, creates a noise-free interference replica, and coherently subtracts it from the received signal. To evaluate the performance limits, we employ the performance metric termed \textit{interference rejection ratio} (IRR), originally proposed in \cite{EB22}, which quantifies the interference canceled. We derive the analytical expressions of IRR as a function of the optimal estimation variances of the signal parameters. Simulation results confirm the accuracy of the analytical expression for both single-carrier and multi-carrier interference signals and demonstrate that the method can substantially suppress the interference at a sufficient interference-to-noise ratio (INR), enabling enhanced detection and extraction of the SOI. We further extend the analysis to the scenario where the SOI is above the noise floor, and confirm the validity of the theoretical IRR expression in this scenario. Lastly, we compare the Demod-Remod technique to other time-domain cancellation methods. The result of the comparison identifies the conditions under which each method is preferred, offering practical guidelines for interference mitigation under different scenarios.
\end{abstract}

\begin{IEEEkeywords}
Interference mitigation, radio frequency interference (RFI), radio astronomy, signal reconstructions.
\end{IEEEkeywords}

\section{Introduction}
In many modern sensing and monitoring applications, such as radio astronomy, passive radar, satellite earth stations, over-the-horizon radar, and low-probability-of-intercept (LPI) radar, the received signal of interest (SOI) is often buried beneath a combination of strong interference and background noise. Unlike conventional radar systems, these systems must operate in environments where interference from external sources may dominate the received SOI by several orders of magnitude. Therefore, active suppression of the RFI is critical to ensure reliable detection and recovery of the SOI. \par
\textit{Interference Mitigation}: Traditional approaches to RFI mitigation, particularly in fields such as radio astronomy and satellite communications, have relied on regulatory protections or hardware-based techniques. For example, in satellite communication systems, Low Earth Orbit (LEO) could potentially create interference to the earth stations for the Geosynchronous Earth Orbit (GEO) satellites. Such interference can be alleviated by constructing protection areas \cite{8633412}, or by introducing nulls in the received pattern in the direction of the interference on a single dish reflector antenna \cite{9400735, 10403767}. However, such methods are increasingly challenged as the electromagnetic environment becomes more complex and congested. This is particularly evident in passive radar systems, which exploit noncooperative illuminators of opportunity (IOs), such as FM, DVB-T, or cellular signals, for target tracking and detection \cite{griffiths2005passive, baker2005passive}. \par
Alternative interference mitigation strategies include spatial techniques such as forming nulls in the direction of the interference or exploring the autocorrelation of the interference through antenna array or multiple geographically distributed receivers \cite{8946890, 10535463, 7849139}. Additionally, adaptive filtering approaches have been widely adopted, where a reference signal is obtained from a dedicated reference channel \cite{7131168, colone2009multistage, palmer2012dvb}. In radio astronomy, similar reference-based filtering techniques have been developed, the authors in \cite{EB22} proposed a filtering method that also uses a reference signal. The result of the comparison between the received signal and the reference signal is used to create an interference estimate, which is subsequently subtracted from the received signal. These methods either require multiple receivers or heavily rely on access to a high-quality reference signal that is highly correlated with the interference observed at the main antenna. When such a reference is weak, delayed, or unavailable, their performance degrades significantly. \par
In these cases, alternative approaches which require no additional receiver include ``short time sinusoidal analysis” (STSA) based cancellation \cite{E20-STSA, li2025parametric}, which uses the fact that most communications and radar signals are comprised of one or more modulated sinusoidal carriers. Estimating the interference reduces to estimating the amplitude, frequency and phase of a sinusoid when the observation window is shorter than the inverse of the interference bandwidth. STSA has modest performance and exhibits relatively high computational complexity, but requires no prior knowledge of the interference signal.  \par
When partial knowledge of the interference structure is available, the authors in \cite{MPM97} present an RFI extraction system, which also relies on parametric estimation and subtraction, for ultrawideband radar. Another promising solution is the Demodulation-Remodulation (Demod-Remod) method. This approach attempts to demodulate the interference signal, reconstruct a clean replica of the interference, and then subtract it from the received signal. Recent studies have shown that it is possible to extract the communications parameters required to successfully demodulate the interference signals and use them as effective references \cite{10549356, EBB00, Nigra+2010, mattingly2024performance, harrington2024challenges, lee2008coherent}. \par
\textit{Contributions}: Existing literature on interference mitigation using the Demod-Remod approach either assumes ideal conditions, such as perfect synchronization, or lacks theoretical analysis of the performance. In this work, we examine the performance limits of the Demod-Remod approach for removing the interference in scenarios where the SOI has a significantly wider bandwidth or weaker power than the interference signal. Under these conditions, the received samples of the SOI can be approximated as independent Gaussian random variables. With only limited prior knowledge of the interference signal structure, the canceler estimates the key parameters—such as amplitude, carrier frequency offset, phase offset, symbol timing offset, modulation type, and symbol values—to reconstruct a clean replica of the interference, which is subsequently subtracted from the received signal. While many works focus on the comparison of the spectrum before and after interference cancellation, we adopt and extend the \textit{interference rejection ratio} (IRR), originally proposed in \cite{EB22}, as a metric to quantify the amount of interference removed. While the IRR has been theoretically analyzed for the filtering-based \cite{EB22} and the STSA-base methods \cite{li2025parametric}, its behavior and analytic characterization under the Demod-Remod framework remains unstudied. In particular, we demonstrate that the IRR depends on the accuracy of the estimated parameters used during the demodulation process. \par
Furthermore, we show that the theoretical expression for IRR remains valid even when the SOI is no longer buried beneath the noise floor. By approximating the SOI power within the interference bandwidth as the additive Gaussian noise, the expression continues to hold with an appropriately adjusted INR. To evaluate performance, we present numerical results using both synthesized and real-world data, comparing the Demod-Remod method with two other time-domain cancellation techniques: the STSA-based method \cite{li2025parametric} and the filtering with a reference signal approach \cite{EB22}. As no single technique is universally optimal, we identify the conditions under which method performs the best, providing practical guidance for interference mitigation under different scenarios. \par
\textit{Organization}: This paper is organized as follows: Section \ref{sec:II} describes the system model and outlines the framework of the Demod-Remod algorithm for removing interference with partial signal knowledge. Section \ref{sec:III} defines the IRR and derives its theoretical bound for both single-carrier and OFDM signals, including an alternative metric of IRR suitable for the real-world collected data. Section \ref{sec:IV} provides numerical results and a comparison to alternative interference mitigation techniques. Subsequently, Section \ref{sec:V} concludes the paper with a summary of findings. 

\section{System Model and Demod-Remod Interference Cancellation Framework}
\label{sec:II}
\begin{figure}[t!]
\includegraphics[width = \linewidth]{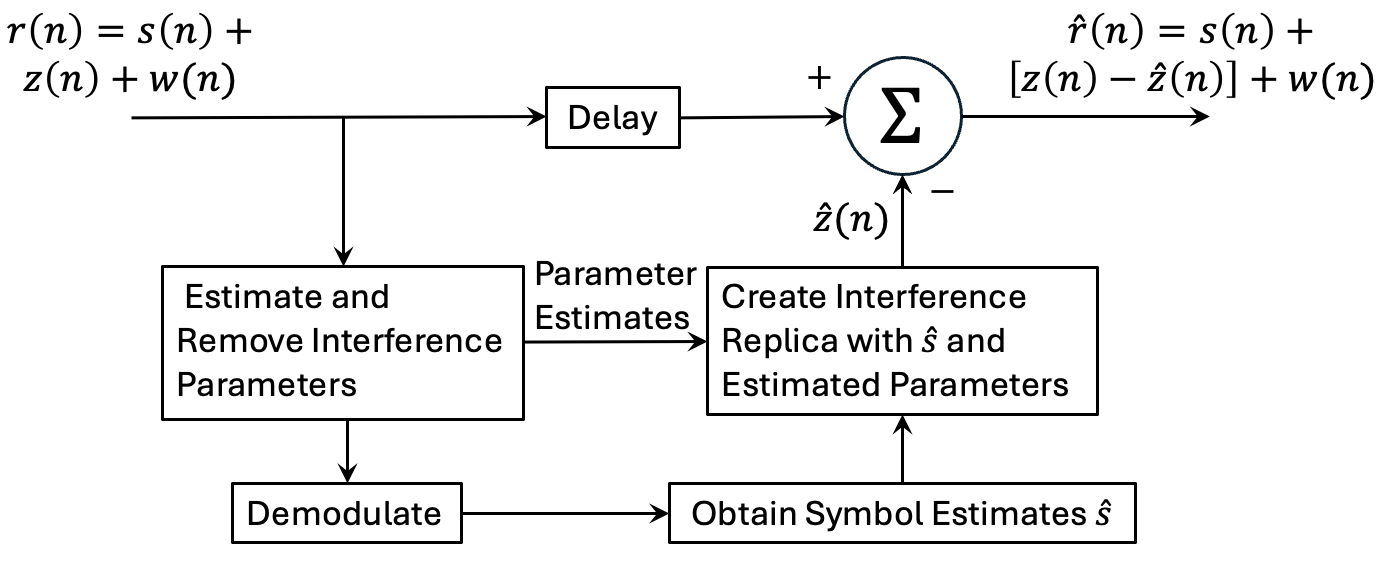}
\caption{Diagram of CTC using the framework of Demod-Remod.}
\label{fig:CTC}
\end{figure}

We begin by expressing the coherent time-domain canceling (CTC) problem under the Demod-Remod framework in discrete-time complex baseband form, as illustrated in Fig.~\ref{fig:CTC}. 
The received signal samples \(r(n)\) are modeled as the sum of the signal of interest (SOI) \(s(n)\) with power \(\sigma_s^2\), the interfering signal \(z(n)\) with power \(\sigma_z^2\), and white Gaussian noise \(w(n)\) with zero mean and variance \(\sigma_n^2\) for \(n = 0, \dots, N-1\). An interference estimate \(\hat{z}(n)\) is synthesized and subsequently subtracted from \(r(n)\) to obtain the residual \(\hat{r}(n) = s(n) + [z(n) - \hat{z}(n)] + w(n)\). The interference signal \(z(n)\) can be written as
\begin{equation}
z(n) = Az_b(n - \epsilon)e^{j(\omega n + \theta)},
\label{eq:z}
\end{equation}
where \(z_b(n)\) is the complex baseband modulated signal, \(A\) is the amplitude which includes the impact of fading and path loss, \(\omega\) is the frequency shift due to the Doppler effect or the mismatch between received center frequency and receiver's local oscillators, \(\theta\) is the phase offset caused by the propagation delay, and \(\epsilon\) is the symbol timing offset in samples. In practice, these parameters are possibly varying slowly with time. However, they can be assumed as static within the estimation window, if this is chosen properly. 

\subsection{Single-Carrier Interference Signal}
If the interference is a single-carrier signal and linearly modulated, the baseband signal \(z_b(n - \epsilon)\) can be expressed as
\begin{equation}
z_b(n - \epsilon) = \sum_{k = -\infty}^{\infty} s_k p(n - \epsilon - kP),
\label{eq:z_b}
\end{equation}
where \(s_k\) is the \(k\)-th transmitted symbol drawn from an unit energy unknown constellation set \(\mathcal{S}\), \(p\) is the pulse shape with unit energy, and \(P\) is the oversampling factor 4for \(s_k\) (i.e., the number of samples per symbol). Assuming that the pulse shape and symbol rate are known, the Demod-Remod method begins by matched filtering the received signal \(r(n)\). Since we assume that any pilot symbols are unknown in the interference \(z(n)\), nondata-aided open-loop estimation structures are necessary to determine the amplitude, frequency, phase, and symbol timing estimates (\(\hat{A}\), \(\hat{\omega}\), \(\hat{\theta}\), \(\hat{\epsilon}\)). \par
Ignoring the impact of the residual symbol timing error, the matched filter output, after removing all the parameter offsets, can be written as
\begin{equation}
y(k) = A/\hat{A} s_ke^{j \left( \Delta_{\omega} P k + \Delta_\theta  \right)} + \nu(k), \quad k = 0, \dots, K-1.
\label{y_k}
\end{equation}
where \(\Delta_{\omega} =  \omega - \hat{\omega}\), \(\Delta_{\theta} =  \theta - \hat{\theta}\), and \(\nu \sim \mathcal{CN}(0, \sigma_{\nu}^2)\) is the complex white Gaussian noise with zero mean and variance \(\sigma_{\nu}^2\). \par
The symbol estimates \(\hat{s}_k\) are obtained by demodulating the sequence \(y(k)\) using the estimated constellation which is blindly determined using a modulation classifier~\cite{6487244}. The interference replica \(\hat{z}(n)\) is synthesized by upsampling the symbol estimates \(\hat{s}_k\), pulse shaping and applying the estimated parameter offsets:
\begin{equation}
\hat{z}(n) = \hat{A} \hat{z}_b(n - \hat{\epsilon})e^{j \left(\hat{\omega} n + \hat{\theta} \right)},
\label{eq:z_hat}
\end{equation}
where 
\begin{equation}
\hat{z}_b(n - \hat{\epsilon}) = \sum_{k = -\infty}^{\infty} \hat{s}_k p(n - \hat{\epsilon} - kP).
\end{equation}
Finally, the interference estimate \(\hat{z}(n)\) is subtracted from the received signal, \(\hat{r}(n) = s(n) + [z(n) - \hat{z}(n)] + w(n)\), ideally completely removing the interference for further processing. 

\subsection{OFDM Interference Signal}
If the interference is an OFDM signal, the complex baseband signal in (\ref{eq:z}) becomes
\begin{equation}
z_b(n - \epsilon) = \sum_{l = -\infty}^{\infty} x_l(n - \epsilon - lP(M + L) ),
\label{eq:z_OFDM}
\end{equation}
where \(x_l(n)\) is the \(l\)-th OFDM symbol and each of which can be expressed as
\begin{equation}
\begin{split}
x_l(n) &= \frac{1}{\sqrt{PM}}\sum\limits_{m=0}^{M-1} s_{m,l} e^{j2\pi (n-PL) \frac{m}{PM}},\\
& \qquad \text{if } \ PL \leq n \leq P(M + L) - 1, \\
&= x(n+PM), \\
& \qquad \text{if} \   0 \leq n < PL,
\label{eq:x_OFDM}
\end{split}
\end{equation}
where \(s_{m,l}\) is the data symbol modulating the \(m\)-th subcarrier of the \(l\)-th OFDM symbol, \(P\) is the oversampling factor, \(M\) is the number of subcarriers, \(L = \kappa M\) is the length of cyclic prefix (CP), \(\kappa\) is the CP ratio, and each OFDM symbol contains \(P(M+L)\) number of samples. After estimating and removing the signal parameter offsets, each received \(l\)-th OFDM symbols becomes
\begin{equation}
\hat{x}_l(n) = A/\hat{A} x_l(n)e^{j \left( \Delta_{\omega} n + \Delta_\theta  \right)} + w(n). 
\end{equation}
By applying Discrete Fourier Transform (DFT) with downsampling for the resulting samples, the decision variable \(y_l(k)\) in each OFDM symbol can be written as
\begin{equation}
y_l(k) = \frac{1}{\sqrt{PM}}\sum_{n = 0}^{PM-1} \hat{x}_l(n) e^{-j 2 \pi k \frac{n}{PM}}, 
\end{equation}
Similarly, the symbol estimates \(\hat{s}_{m,l}\) are obtained by demodulating, i.e., demapping the sequence \(y_l(k)\) to the estimated constellation, and the interference estimate is created using (\ref{eq:z_hat}), (\ref{eq:z_OFDM}) and (\ref{eq:x_OFDM}) with the parameter offsets applied. \par
While oversampling is not commonly used in multi-carrier systems primarily because it offers limited benefit when using low-complexity receivers that assume orthogonal subcarriers, we still consider oversampled OFDM signals in our analysis. This is motivated by two factors. First, oversampled OFDM signal appears in various practical applications, such as blind frequency estimation or peak-to-average power ratio (PAPR) reduction \cite{1306655, 6104194, 1040728}. Second, since oversampling and subsequent pulse shaping are commonly used in single-carrier systems, including them in the OFDM case allows for a more equitable performance comparison. It is worth noting that the theoretical framework developed for the oversampled OFDM case in the following remains valid for the non-oversampled case, provided that the relevant parameters are appropriately scaled. \par

\section{Theoretical Limits of Performance} 
\label{sec:III}

We now address the theoretical limits of interference cancellation when using the Demod-Remod cancelling algorithm.  

\subsection{Definition of the IRR Metric}
To evaluate the theoretical performance limits of the Demod-Remod interference cancellation algorithm, we calculate the \textit{Interference Rejection Ratio} (IRR), which was initially proposed in \cite{EB22}. IRR measures how effectively the interference is removed and is defined as the ratio of the power of \(z(n)\) to the power of the same signal as it appears in the output; i.e.,  \(z(n) - \hat{z}(n)\). 
Using previously established notation, this can be written as
\begin{equation}
\begin{split}
\mathrm{IRR} &= \frac{\frac{1}{N} \sum\limits_{n=n_0}^{n_0+N-1} \left |z(n) \right |^2}
                {\frac{1}{N} \sum\limits_{n=n_0}^{n_0+N-1} \left |z(n)-\hat{z}(n) \right |^2} 
\label{eIRR2Def}                    
\end{split}
\end{equation}
Since $N$ is finite and the estimated waveform parameters are essentially random variables, IRR is itself a random variable. However, directly averaging IRR values across trials (i.e., \(\mathbb{E}[\mathrm{IRR}]\)) is problematic due to the heavy-tailed nature of its distribution, where a few large values can dominate the average. This problem is solved in \cite{li2025parametric} by taking the expectation of the numerator and denominator for each $N$-sample estimation interval, leading to a new metric:
\begin{equation}
\overline{\mbox{IRR}} = \frac{\mathbb{E}\left[\frac{1}{N} \sum\limits_{n=n_0}^{n_0+N-1} \left |z(n) \right |^2\right]} 
                {\mathbb{E}\left[\frac{1}{N} \sum\limits_{n=n_0}^{n_0+N-1} \left |z(n)-\hat{z}(n) \right |^2\right]},
\label{eIRRBarDef} 
\end{equation}
where the expectation is taken over the parameter estimates. Note that $\overline{\mbox{IRR}}$ has the desirable property that it converges with increasing number of trials used to compute the expectations in the numerator and denominator. Additionally, the definition of $\overline{\mbox{IRR}}$ facilitates analysis yielding to a closed-form expression in terms of the relevant parameters INR and $N$. This is accomplished in the next subsection. \par
Before proceeding, we note that in the special case of Parametric Estimation and Subtraction (PES)-based canceling of a modulated signal, the numerator of (\ref{eIRRBarDef}) is simply the power of the interference signal \(\sigma_z^2\), which is equal to \(\frac{A^2}{P}\).  Furthermore, we define the denominator of (\ref{eIRRBarDef}) as
\begin{equation}
   \xi(N) =  \mathbb{E}\left[ \frac{1}{N} \sum\limits_{n=n_0}^{n_0+N-1} \left |z(n)-\hat{z}(n) \right |^2 \right],
\end{equation}
which represents the mean squared error of the interference estimate with an estimation block size \(N\). Thus:
\begin{equation}
\overline{\mbox{IRR}} = \frac{A^2}{\xi(N) P}.
\label{eq:IRR_def}
\end{equation}

\subsection{Analysis Using the \texorpdfstring{$\overline{\mathrm{IRR}}$}{IRR} Metric for Single-Carrier Interference } 
In Appendix \ref{app:A}, we show that by letting \(n_0=-(N-1)/2\), \(\xi(N)\) can be evaluated as
\begin{equation}
\begin{split}
\xi(N) &= \frac{2A^2}{P} + \frac{\sigma_A^2}{P} - \gamma \frac{2A^2}{P N}  e^{-\frac{\sigma_{\theta}^2}{2}} \sum_{n=-(N-1)/2}^{(N-1)/2} e^{-\frac{n^2}{2}\sigma_{\omega}^2}, \\
&\approx \frac{2A^2}{P} + \frac{\sigma_A^2}{P} - \gamma  e^{-\frac{\sigma_{\theta}^2}{2}} \frac{2\sqrt{2\pi} A^2}{ P N \sigma_\omega} \mathrm{ erf} \left(\frac{N\sigma_\omega}{2\sqrt{2}} \right),
\label{eq:xi}
\end{split}
\end{equation}
where \(\sigma_A^2\), \(\sigma_{\theta}^2\) and \(\sigma_{\omega}^2\) are the estimation error variances for the amplitude, phase and frequency. The parameter \(\gamma \in (0,1]\) accounts for the errors in the modulation classification, symbol timing and demodulation for the single-carrier signal, and is defined as
\begin{equation}
\gamma = \left( 1 - \frac{P_c d_1^2 P_s + (1 - P_c)d_2^2}{2} \right)\left(1 - \frac{\sigma_{\epsilon}^2 E_p'}{2}  \right),
\label{eq:gamma}
\end{equation}
where \(P_c\) is the probability of correct modulation classification, \(P_s\) is the symbol error rate (SER), \(d_1\) is the average error distance between \(s_k\) and \(\hat{s}_k\), which depends on the modulation scheme. \(d_{2}\) is the average distance between the original symbols and their closest symbols from the other modulation schemes being classified. \(\sigma_{\epsilon}^2\) is the estimation variance for the symbol timing \(\epsilon\) and \(E_p' = \sum_n |p'(n)|^2 = P^2 \sum_{n} |p(n) - p(n - 1)|^2\) is the energy of the differentiated pulse. Inserting (\ref{eq:gamma}) in (\ref{eq:xi}) and then (\ref{eq:xi}) in (\ref{eq:IRR_def}), \(\overline{\mathrm{IRR}}\) is approximated as
\begin{equation}
\begin{split}
\overline{\mathrm{IRR}} \approx \Biggl( 2 + \frac{\sigma_A^2}{A^2} - \gamma  e^{-\frac{\sigma_{\theta}^2}{2}} \frac{2\sqrt{2\pi}}{ N \sigma_\omega} \mathrm{ erf} \left(\frac{N\sigma_\omega}{2\sqrt{2}} \right) \Biggl)^{-1}.
\label{IRR_bar}
\end{split}
\end{equation} \par
We notice that in the absence of symbol error, classification error and symbol timing error, (\ref{IRR_bar}) reduces to the IRR for the STSA-based canceler, which is presented in equation (19) of \cite{li2025parametric}. However, \(\overline{\mathrm{IRR}}\) for the STSA-based algorithm equals the theoretical performance only for sinusoids with constant parameters and tends to saturate with increasing INR when applied to modulated signals, as shown in Fig. 6 of \cite{li2025parametric}. In contrast, since the Demod-Remod approach leverages partial knowledge of the signal, i.e., pulse shape and symbol rate, it has superior performance compared to the STSA-based canceler, which will be further demonstrated in section IV. Note that the derived \(\overline{\mathrm{IRR}}\) expression can be modified to accommodate other channel conditions. However, we focus on the additive white Gaussian noise (AWGN) channel scenario for analytical tractability and leave other system setups as potential future work. \par
In the presence of SOI, the accuracy of interference parameter estimation depends on the power and bandwidth of the SOI occupied in the interference bandwidth. When the SOI is buried beneath the noise floor or its bandwidth is sufficiently wider than that of the interference, the estimation process can be effectively modeled as parameter estimation in an AWGN channel. If the SOI has a strongly negative signal-to-noise ratio (SNR), the effective INR is given by \(\mathrm{INR}_{\mathrm{eff}} = \mathrm{INR} = \frac{A^2}{P \sigma_n^2}\). Conversely, when the SOI power is comparable to the interference power, but the bandwidth of the SOI is larger than that of the interference, the interference estimation is still well-approximated by assuming the estimation is done in an AWGN channel, but with an elevated noise floor due to the contribution of the SOI power within the interference bandwidth. This approximation holds because the SOI, being wideband relative to the interference, is approximately uncorrelated from sample to sample and noise-like within the interference bandwidth, thus behaving similarly to white Gaussian noise. In this case, the effective INR becomes 
\begin{equation}
\mathrm{INR}_{\mathrm{eff}} \approx \frac{A^2}{P(\sigma_n^2 + \alpha \sigma_s^2)},
\end{equation}
where \(\sigma_s^2\) is the power of the SOI, and \(\alpha\) represents the fractional bandwidth of the SOI that overlaps with the interference. The effective INR can also be interpreted as the ``interference-to-SOI-plus-noise ratio". Therefore, for a single-carrier signal in white Gaussian noise, the variances of the optimal (i.e., minimum variance unbiased) estimators for amplitude, frequency, and phase are lower-bounded by the Cramér–Rao Lower Bound (CRLB), assuming independence among estimation errors~\cite{Kay93}
\begin{equation}
\begin{split}
\sigma_A^2 &\geq \frac{\sigma_z^2}{2 \ \mathrm{INR}_{\mathrm{eff}}  N}, \\
\sigma_{\omega}^2 &\geq \frac{6}{\mathrm{INR}_{\mathrm{eff}}  N (N^2-1)}, \\
\sigma_{\theta}^2 &\geq \frac{1}{2 \ \mathrm{INR}_{\mathrm{eff}}  N},
\label{eq:crlb}
\end{split}
\end{equation}
On the other hand, the CRLB of the symbol timing when all the rest of the parameters are unknown is lower bounded by the CRLB when the data symbols are known. The CRLB of \(\hat{\epsilon}\) for joint estimation of \(\hat{\omega}\), \(\hat{\theta}\) and \(\hat{\epsilon}\) with known symbols can be written as \cite{729398}
\begin{equation}
\sigma_{\epsilon}^2 \geq \frac{1}{2 \ \mathrm{INR}_{\mathrm{eff}} N E_p'}.
\label{eq:ep_crlb}
\end{equation}
Substituting these into (\ref{IRR_bar}), we get
\begin{equation}
\begin{split}
\overline{\mathrm{IRR}} &\approx 2 + \frac{1}{2 \  \mathrm{INR}_{\mathrm{eff}}  N} - \gamma  e^{-\frac{1}{4  \mathrm{INR}_{\mathrm{eff}}  N}} \\
&\quad \ \cdot \sqrt{\frac{4 \ \mathrm{INR}_{\mathrm{eff}}  N \pi}{3}} \mathrm{erf} \left( \sqrt{\frac{3}{4 \ \mathrm{INR}_{\mathrm{eff}} N}}  \right), 
\end{split}
\end{equation}
and by substituting (\ref{eq:ep_crlb}) into (\ref{eq:gamma})
\begin{equation}
\gamma = \left( 1 - \frac{P_c d_1^2 P_s + (1 - P_c)d_2^2}{2} \right)\left(1 - \frac{1}{4 \ \mathrm{INR}_{\mathrm{eff}} N}  \right).
\label{eq:gamma_sc}
\end{equation}
Here, \(\gamma\) reflects the impact of SER, modulation classification accuracy, and symbol timing error. Since the amplitude, frequency, and phase estimation errors dominate the \(\overline{\mathrm{IRR}}\) at low INR and have negligible impact on the SER at high INR, we assume \(P_s\) follows the theoretical SER in an AWGN channel for each modulation scheme, which also serves as a lower bound for the true SER. Further, the distances \(d_1\) and \(d_2\) depend on the constellation of the interference and the possible results of the modulation classification, respectively. The probability of correct classification \(P_c\) is obtained empirically and depends on the specific classifier used. 

\subsection{Analysis Using the \texorpdfstring{$\overline{\mathrm{IRR}}$}{IRR} Metric for OFDM Interference }
For OFDM signals, the expression for \(\overline{\mathrm{IRR}}\) remains the same as that for the single-carrier case given in (\ref{IRR_bar}), except for the term \(\gamma\), which becomes
\begin{equation}
\gamma = \left( 1 - \frac{P_c d_1^2 P_s + (1 - P_c)d_2^2}{2} \right) \frac{\mathrm{erf} \left(\sqrt{2} \pi \sigma_{\epsilon} \right)}{2\sqrt{2 \pi } \sigma_{\epsilon}},
\label{eq:gamma_ofdm}
\end{equation}
and is derived in Appendix~\ref{app:B}. On the other hand, the CRLB of frequency estimation with an OFDM signal in an AWGN channel, as derived in \cite{moose1994technique}, is given by
\begin{equation}
\begin{split}
\sigma_{\omega}^2 (\mathrm{OFDM}) &\geq \frac{1}{\mathrm{INR}_{\mathrm{eff}} LM^2 P^3} = \frac{(1 + \kappa)^3}{\mathrm{INR}_{\mathrm{eff}} \kappa N^3}.
\label{eq:freq_crlb_ofdm}
\end{split}
\end{equation}
Inserting (\ref{eq:freq_crlb_ofdm}) in (\ref{IRR_bar}) and using the same CRLBs for the amplitude and phase as in the single-carrier case as shown in (\ref{eq:crlb}), the corresponding \(\overline{\mathrm{IRR}}\) for the OFDM signal becomes
\begin{equation}
\begin{split}
\overline{\mathrm{IRR}} &\approx 2 + \frac{1}{2 \  \mathrm{INR}_{\mathrm{eff}} N} - \gamma  e^{-\frac{1}{4  \mathrm{INR}_{\mathrm{eff}} N}} \\
&\quad \ \cdot \frac{\sqrt{8 \ \mathrm{INR}_{\mathrm{eff}} \kappa N \pi}}{1+\kappa} \mathrm{erf} \left(\frac{1 + \kappa}{\sqrt{8 \ \mathrm{INR}_{\mathrm{eff}}\kappa N}} \right).
\end{split}
\end{equation}
Although an OFDM signal is more sensitive to frequency offset than a single-carrier signal, we find that the SER of OFDM after frequency offset correction is nearly equivalent to that of a single-carrier signal in an AWGN channel. Therefore, we assume that the SER \(P_s\) for OFDM uses the same expression as for the single-carrier case. The rest of the parameters in (\ref{eq:gamma_ofdm}) are computed using the same manner as in (\ref{eq:gamma_sc}) with the corresponding estimator.

\subsection{Measuring IRR from Collected Data}
When evaluating interference cancellation performance using real-world collected data, we typically do not have a access to a noise-free version of the interference signal \(z(n)\). Therefore, \(\overline{\mathrm{IRR}}\) cannot be directly computed using (\ref{eIRRBarDef}). To address this limitation, we adopt an alternative metric \(\overline{\mathrm{IRR}}_c\), originally proposed in Section 4 of \cite{li2025parametric}, which is specifically designed to quantify the amount of interference removed in practical scenarios where only noisy measurements are available. Referring to the definition in \cite{li2025parametric}, \(\overline{\mathrm{IRR}}_c\) is given by
\begin{equation}
\overline{\mathrm{IRR}}_c = \frac{\mathbb{E}\left[ \frac{1}{N} \sum\limits_{n = n_0}^{n_0+N-1} |z(n) + w(n)|^2\right]}{\mathbb{E}\left[ \frac{1}{N} \sum\limits_{n = n_0}^{n_0+N-1} |z(n) + w(n) - \hat{z}(n)|^2\right]}.
\end{equation}
This ratio is both meaningful and straightforward to compute in practical applications. In scenarios such as radio astronomy, where the SOI is buried beneath the noise floor, the received signal can be effectively modeled as a combination of interference and noise. In contrast, for radar applications where the SOI is not necessarily buried in noise, interference-only data can be obtained by simply suspending SOI transmission during a calibration period when examining the performance of the canceler. By comparing the total received power before and after the interference cancellation, under conditions where the SOI is absent, this ratio captures the relative amount of interference energy suppressed by the canceler. \par
Assuming that the interference estimate is uncorrelated with the noise, i.e., \(\mathbb{E}[\hat{z}(n)w(n)] \approx 0 \), \(\overline{\mathrm{IRR}}_c\) for the Demod-Remod approach can be approximated as
\begin{equation}
\begin{split}
\overline{\mathrm{IRR}}_c & \approx \frac{A^2 + \sigma_n^2}{2A^2 + \sigma_A^2 - \gamma \frac{2A^2}{N}  e^{-\frac{\sigma_{\theta}^2}{2}} \sum\limits_{n=-\frac{N-1}{2}}^{\frac{N-1}{2}} e^{-\frac{n^2}{2}\sigma_{\omega}^2} + \sigma_n^2}, \\
&\approx \frac{A^2 + \sigma_n^2}{2A^2 + \sigma_A^2 - \gamma  e^{-\frac{\sigma_{\theta}^2}{2}} \frac{2\sqrt{2\pi} A^2}{ N \sigma_\omega} \mathrm{ erf} \left(\frac{N\sigma_\omega}{2\sqrt{2}} \right) + \sigma_n^2},
\end{split}
\end{equation}
where the key distinction from \(\overline{\mathrm{IRR}}\) is that the noise variance is included in both the numerator and denominator. Notably, when the interference signal \(z(n)\) is perfectly canceled, \(\overline{\mathrm{IRR}}_c\) reduces to \(\mathrm{INR} + 1\). On the other hand, to prove the validity of \(\overline{\mathrm{IRR}}_c\) in quantifying the \(\overline{\mathrm{IRR}}\) for the collected data, we can show that
\begin{equation}
\begin{split}
\lim_{\substack{\sigma_A, \sigma_{\theta}, \sigma_{\omega}, \sigma_{\epsilon}, P_s \rightarrow 0\\ P_c \rightarrow 1}} \overline{\mathrm{IRR}}_c &\approx \frac{A^2 + \sigma_n^2}{2A^2 - \frac{2A^2}{N}N + \sigma_n^2}, \\
&= \frac{A^2 + \sigma_n^2}{\sigma_n^2} = \mathrm{INR} + 1. 
\end{split}
\end{equation}
Since perfect estimation provides the same value of \(\overline{\mathrm{IRR}}_c\) as zero interference, we conclude that \(\overline{\mathrm{IRR}}_c\) is a valid and meaningful surrogate for \(\overline{\mathrm{IRR}}\) when analyzing cancellation performance using collected data, especially in scenarios where direct access to the clean interference signal is not available.

\section{Simulation Results}
\label{sec:IV}

\begin{figure}[t!]
\includegraphics[width = \linewidth]{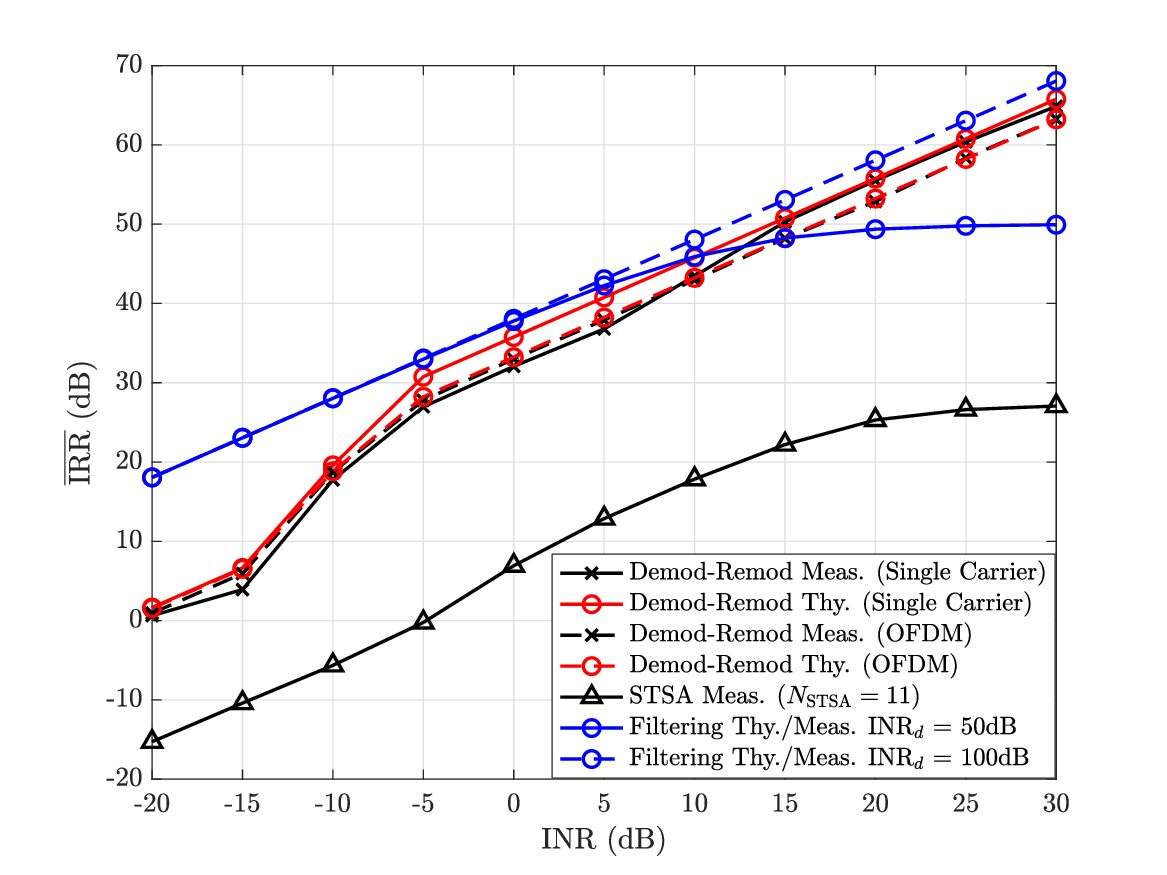}
\caption{INR vs. \(\overline{\mathrm{IRR}}\) for single-carrier and OFDM interference (both with QPSK modulation) with the LFM signal as SOI, where \(\sigma_s^2 \ll \sigma_z^2\). INR$_d$ denotes the INR of the reference signal for the filtering case, and \(N_{\mathrm{STSA}}\) is the estimation window for the STSA-based technique. Number of symbols \(K = 69\) corresponds to \(N \approx 6000\) samples after pulse shaping. The theoretical and measured performance of the filtering technique are shown as a single curve, as the experimental results consistently match the theoretical bound. }
\label{fig:IRR}
\end{figure}

\begin{figure}[t!]
\includegraphics[width = \linewidth]{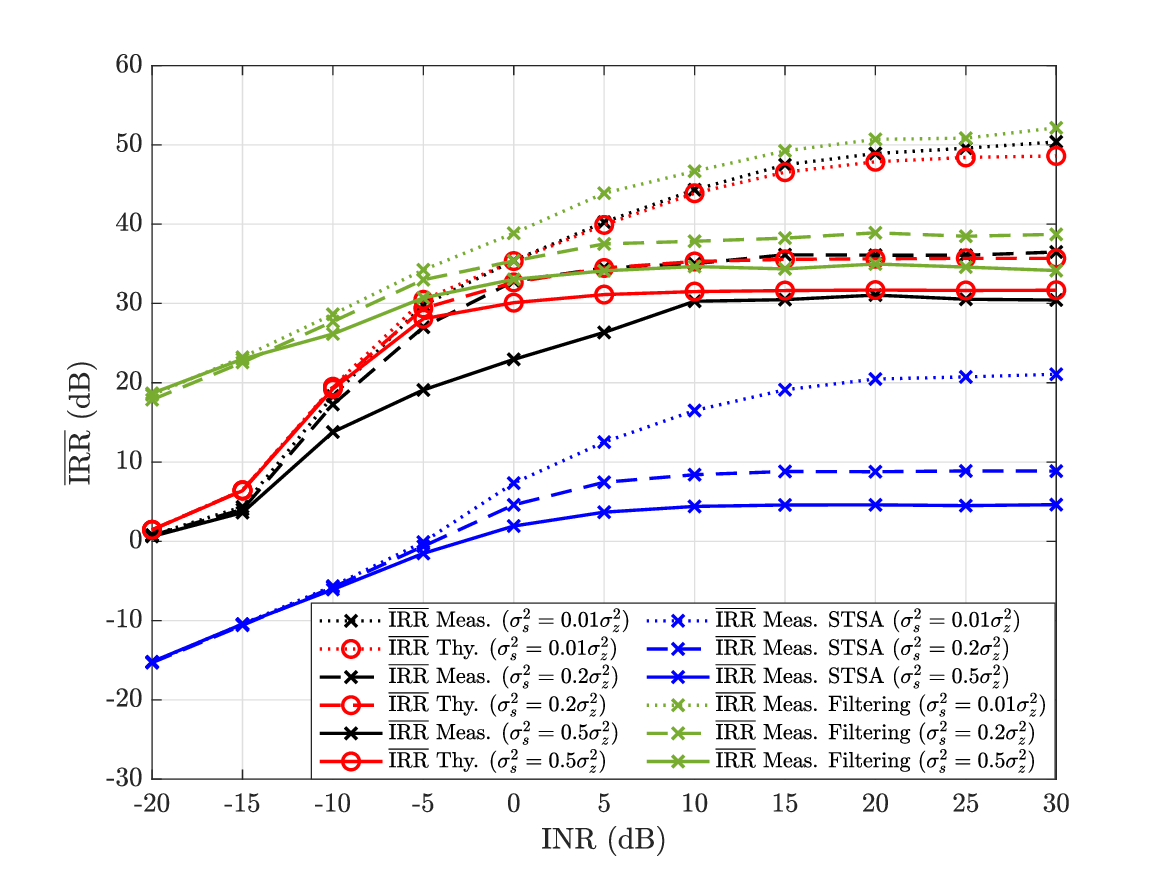}
\caption{INR vs. \(\overline{\mathrm{IRR}}\) for QPSK interference with the LFM signal as SOI at different power levels, for filtering technique \(\mathrm{INR}_d = 100\)dB, and for STSA \(N_{\mathrm{STSA}} = 11\). }
\label{fig:IRR_SOI}
\end{figure}

To demonstrate the performance of Demod-Remod-based cancellation technique, we present numerical results in this section. Specifically, we use the payload portion of the Iridium simplex downlink signal as the interference signal. The associated signal is a QPSK modulated signal with a channel burst symbol rate of 25kbps and is located in L-band between 1626.0 and 1626.5 MHz. The symbols are pulse-shaped using a root-raised cosine filter with a roll-off factor 0.4, resulting in an overall bandwidth of 31.5 kHz \cite{Martincomm}. To ensure consistency with our collected Iridium data and the simulation setups used in \cite{EB22} and~\cite{li2025parametric}, the symbols in the simulations use an upsampling factor of \(P = 82\), which corresponds to a sampling rate of \(f_s = 2.048\)MHz. The pulse shape spans 21 symbols. The SOI is modeled as a linearly frequency modulated (LFM) chirp signal with a signal structure:
\begin{equation}
s(n) = A_s e^{j 2\pi \left(f_0 n + \frac{c}{2}n^2 \right)},
\end{equation}
where \(A_s\) is the scaling factor for the SOI, \(f_0\) is the start normalized frequency, and \(c\) denotes the chirp rate.  \par

\begin{figure*}
\includegraphics[width=\textwidth]{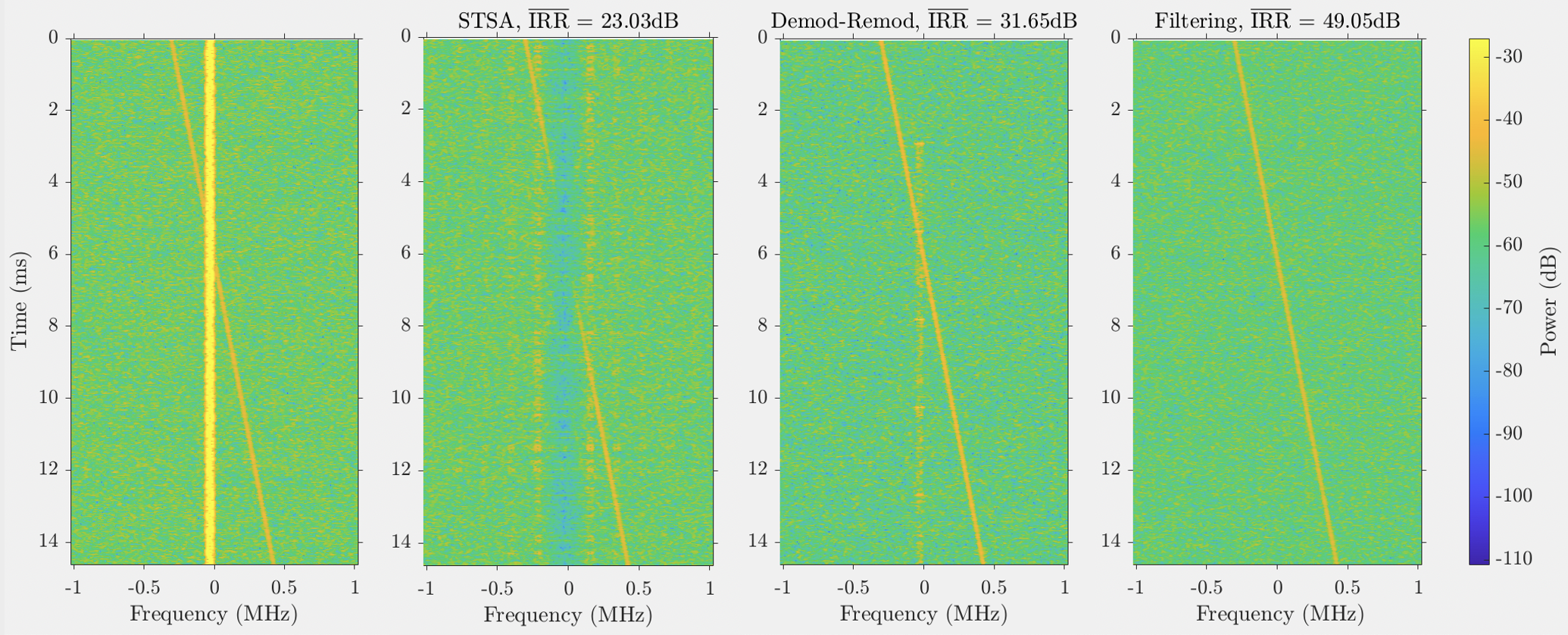}
\caption{Time-Frequency plots for LFM radar signal using different cancellation strategies. Original spectrum (far left), STSA (middle left), Demod-Remod (middle right), Filtering (far right). The SOI power is 6dB lower than the interference power. INR = 20dB and INR$_d$ = 50dB for the filtering technique. \(N_{\mathrm{STSA}} = 33\). }
\label{fig:dynamic}
\end{figure*}

In the Demod-Remod algorithm, frequency offset and symbol timing are blindly estimated using the Cyclostationary approach described in \cite{662646} for a single-carrier signal, or a maximum likelihood (ML) estimator as presented in \cite{599949} for an OFDM signal. The amplitude and phase offsets are estimated using an ML approach after obtaining the frequency offset. For modulation classification, a Kolmogorov-Smirnov (K-S) classifier is employed to distinguish among BPSK, QPSK, 8-PSK, 16-QAM, and 64-QAM signals \cite{5555884}. On the other hand, the simulation procedures for STSA-based cancellation and the filtering technique for comparison are specified in Section~5 of~\cite{li2025parametric}. \par
We begin by evaluating the performance of the Demod–Remod algorithm and validating the accuracy of the theoretical expression of \(\overline{\mathrm{IRR}}\) when the SOI is below the noise floor. To account for the time-varying nature of real satellite signals, we restrict the parameter estimation window to approximately 6000 samples (i.e., 3ms at the 2.048MHz sampling rate). Note that the K-S classifier is used on a longer sequence of data (i.e., the whole 18ms of an Iridium pulse) to ensure the classification accuracy at high INR. On the other hand, the scenario for the OFDM interference is included to illustrate the performance of the proposed method under different interference types. For fairness, we make both single-carrier and OFDM systems transmit the same number of symbols by letting \(K = M + L\). Specifically, we use \(K = 69\) symbols for the single-carrier signal and \(M = 64\) for OFDM with a cyclic prefix length \(L = 5\) (\(\kappa \approx 7\)\%). \par
In Fig.~\ref{fig:IRR}, the performance of the Demod-Remod technique exhibits a noticeable qualitative change around the \(\mathrm{INR} =-5\)dB point. Empirical results show that the SER becomes negligible when the INR exceeds \(-5\)dB, which corresponds to an \(E_s/N_0\) of approximately 14dB given the oversampling rate \(P = 82\). As a result, the performance of the Demod-Remod method is impacted by both the SER and parameter estimation errors at low INR, but primarily affected by the estimation error at high INR where the SER becomes negligible. Moreover, due to the compromised frequency estimation accuracy of OFDM, its theoretical IRR is slightly lower than that of the single-carrier signal. Fig. \ref{fig:IRR} also includes a comparison between the Demod-Remod, STSA, and the filtering-based methods for the single-carrier signal. We can see that the \(\overline{\mathrm{IRR}}\) for the STSA saturates for large values of INR due to the varying amplitude, phase and frequency within each estimation window. Therefore, we choose \(N_{\mathrm{STSA}}=11\) as the estimation block size since larger block size may degrade the performance for STSA. On the other hand, for the filtering technique with training length equal to \(N\), the cancellation performance outperforms the other two when a reference signal is available with high INR. However, this advantage at high INR levels diminishes if INR$_d$ not high enough (i.e., \(\mathrm{INR}_d = 50\)dB, as shown in Fig.~\ref{fig:IRR}), and the performance continues to degrade at lower INR levels as the INR$_d$ decreases further. \par
Since Fig.~\ref{fig:IRR} shows similar performance for both single-carrier and OFDM interference, the remaining simulations focus on the single-carrier case. In Fig.~\ref{fig:IRR_SOI}, we evaluate the performance of the Demod–Remod algorithm in the presence of the SOI that is no longer buried beneath the noise floor. The SOI power is varied relative to the interference, and the interference is configured with a starting frequency of \(f_0 = 0.05\), and a chirp rate of \(c = 4\times 10^{-5}\), resulting in a normalized bandwidth of approximately 0.24. The signal-to-interference-ratio (SIR) varies from \(-20\)dB (\(\sigma_s^2 = 0.01\sigma_z^2\)) to \(-3\)dB (\(\sigma_s^2 = 0.5\sigma_z^2\)). The interference occupies 7\% of the SOI's bandwidth. As shown in Fig.~\ref{fig:IRR_SOI}, the theoretical expression for \(\overline{\mathrm{IRR}}\) remains accurate for low SIR. However, at moderate INR, the measured \(\overline{\mathrm{IRR}}\) begins to deviate from the theoretical bound at a high SIR. Empirical results indicate that this deviation arises from frequency and phase estimation errors, which fail to achieve their respective CRLBs due to the non-Gaussianity of SOI and noise samples. Additionally, we observe that as the SOI power \(\sigma_s^2\) increases, \(\overline{\mathrm{IRR}}\) saturates at a lower level due to the reduced INR$_\text{eff}$. \par
\begin{figure}[t!]
\includegraphics[width = \linewidth]{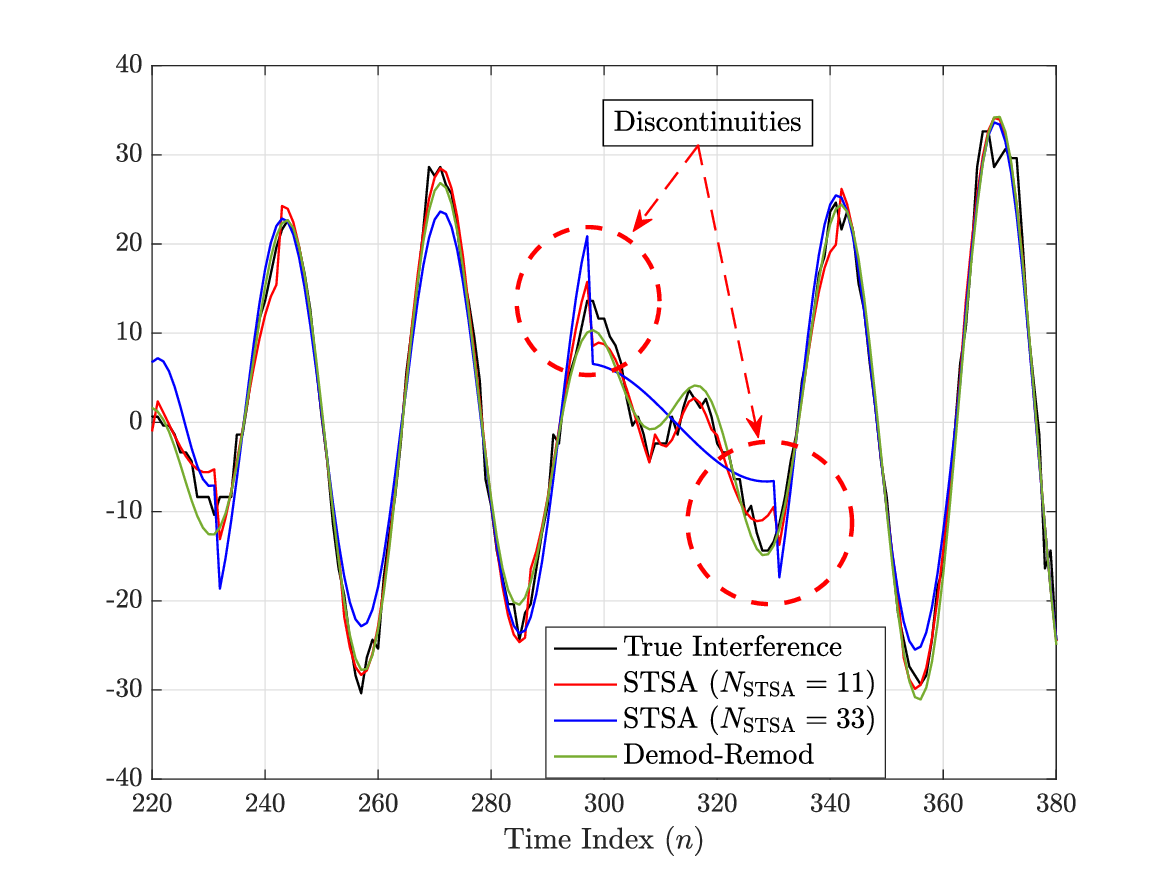}
\caption{Real part of the interference signal and its estimates generated by different cancellation techniques. Discontinuities introduced by the STSA-based method are highlighted with red dashed circles. }
\label{fig:time}
\end{figure}

Fig.~\ref{fig:dynamic} shows an example time–frequency plot before and after the cancellation of the interference using various techniques. The interference duration is set to 15ms, and the estimation window size for the Demod-Remod approach is fixed at 6000 samples. While the STSA method is effective in suppressing the interference, it introduces noticeable distortion within the interference bandwidth and minor spectral distortion to the SOI outside. The in-band distortion arises from the “scooping” effect, caused by spurious correlations with noise; i.e., the parametric estimator occasionally detects sinusoidal components within the noise \cite{li2025parametric}. Meanwhile, the out-of-band spectral distortion arises from the discontinuities between adjacent estimation windows, as illustrated in Fig.~\ref{fig:time}. In contrast, the Demod-Remod algorithm does not fully cancel the interference but introduces significantly less distortion to the SOI. Lastly, the filtering approach yields the best performance, achieving almost complete interference cancellation with minimal impact on the SOI. This is primarily due to the availability of a higher INR reference copy of the interference signal from a reference antenna. \par

\begin{figure}[t!]
\includegraphics[width = \linewidth]{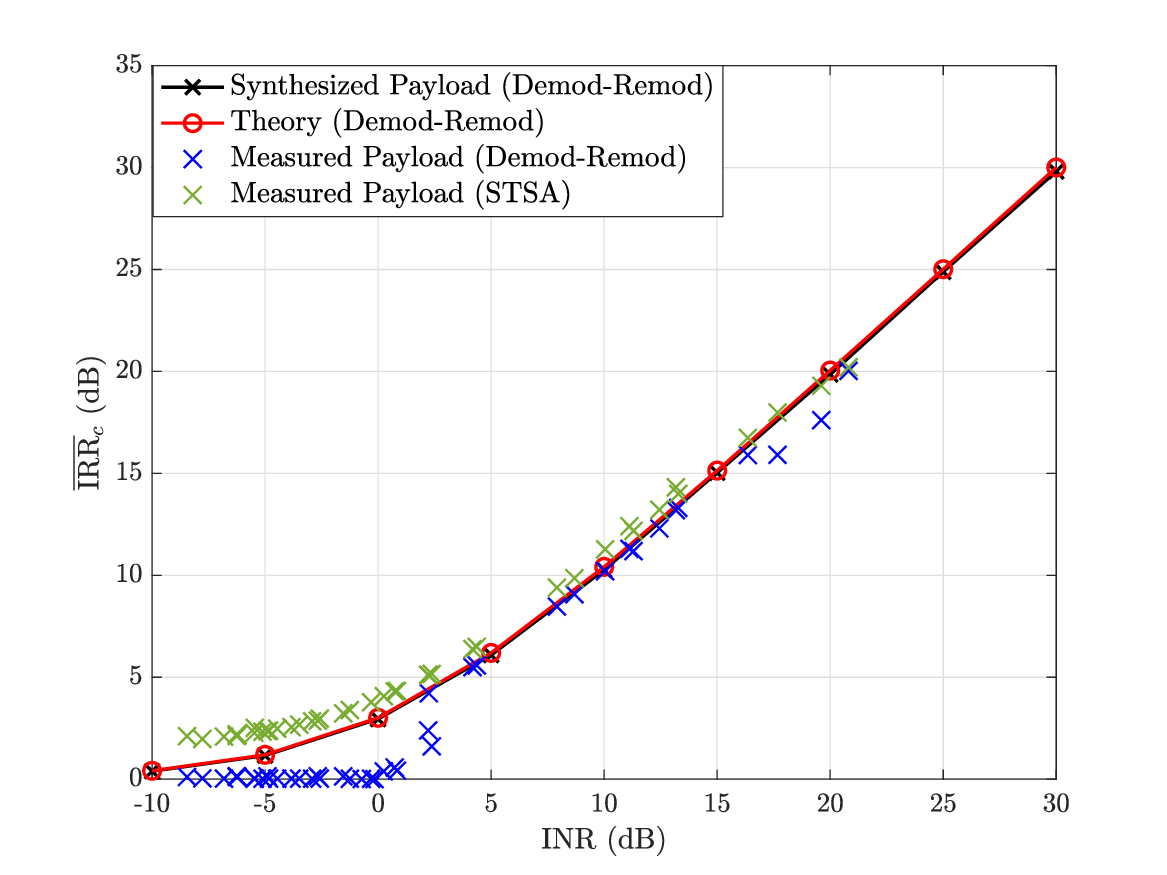}
\caption{INR vs. \(\overline{\mathrm{IRR}}_c\) for synthesized QPSK interference, and \(\overline{\mathrm{IRR}}_c\) samples of the Demod-Remod and STSA approaches for the collected Iridium signal. }
\label{fig:IRRc}
\end{figure}

We further evaluate the performance of the Demod-Remod approach using real-world Iridium data, where the methods for capturing the data and calculating the INR are described in the Section 5.3 of \cite{li2025parametric}. The estimation window length is again set to 6000 samples to account for time-varying frequency offsets induced by Doppler shifts. Since the true interference signal is not available for the collected data, we can only measure \(\overline{\mathrm{IRR}}_c\). Fig.~\ref{fig:IRRc} presents the empirical results of \(\overline{\mathrm{IRR}}_c\) for both the collected data and the synthesized payload used for comparison. Each blue ``\(\times\)" represents a measured \(\overline{\mathrm{IRR}}_c\) using the Demod-Remod method, while each green ``\(\times\)" represents a measured \(\overline{\mathrm{IRR}}_c\) using the STSA method. The figure shows that the Demod-Remod algorithm achieves cancellation performance comparable to STSA at moderate to high INRs. However, at lower INR range, its performance degrades due to larger parameter estimation errors and higher SER. \par
\begin{figure}[t!]
\includegraphics[width = \linewidth]{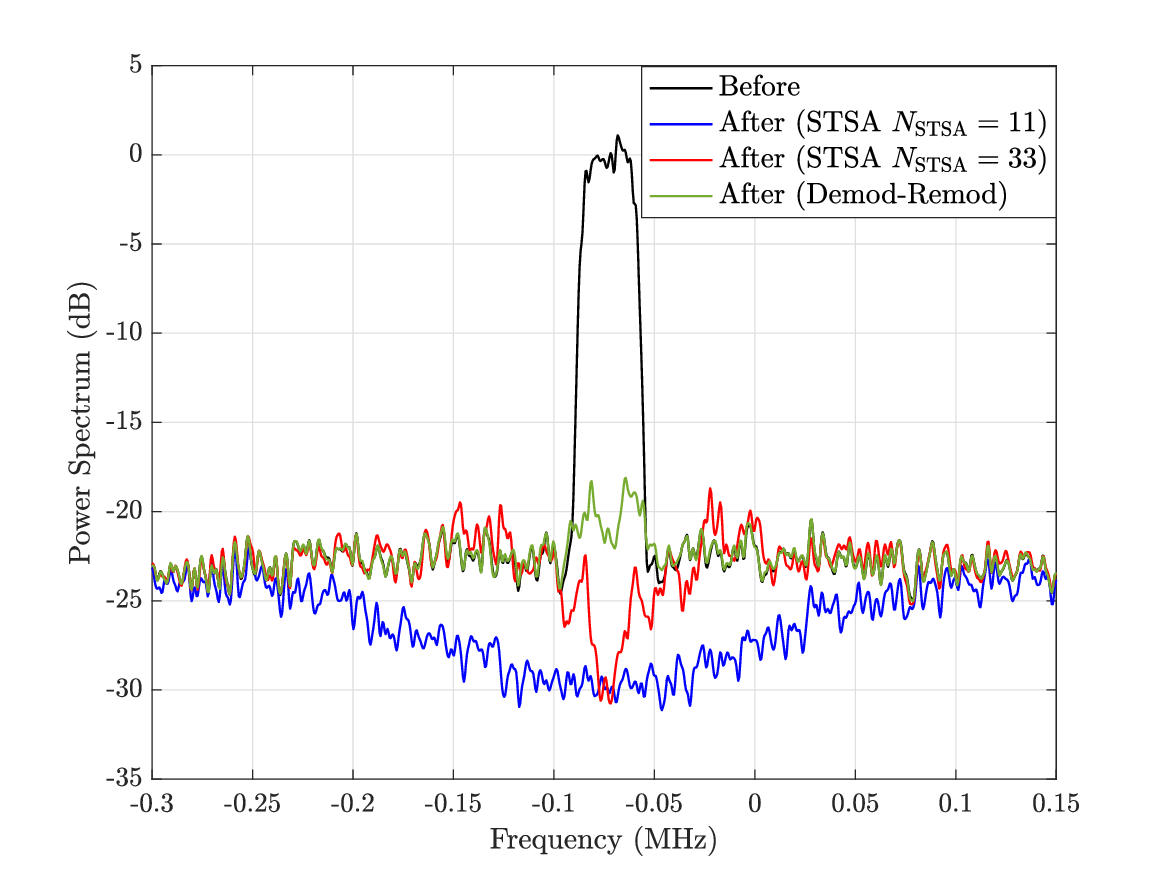}
\caption{Power spectrum for QPSK modulated signal of collected Iridium data before and after coherent subtraction with STSA and Demod-Remod approaches. \(\mathrm{INR} \approx 4.17\)dB, \(\overline{\mathrm{IRR}}_c = 5.5\)dB for STSA with \(N_{\mathrm{STSA}} = 11\), \(\overline{\mathrm{IRR}}_c = 3.96\)dB for STSA with \(N_{\mathrm{STSA}} = 33\), and \(\overline{\mathrm{IRR}}_c = 5.49\)dB for the Demod-Remod. }
\label{fig:psplow}
\end{figure}
\begin{figure}[t!]
\includegraphics[width = \linewidth]{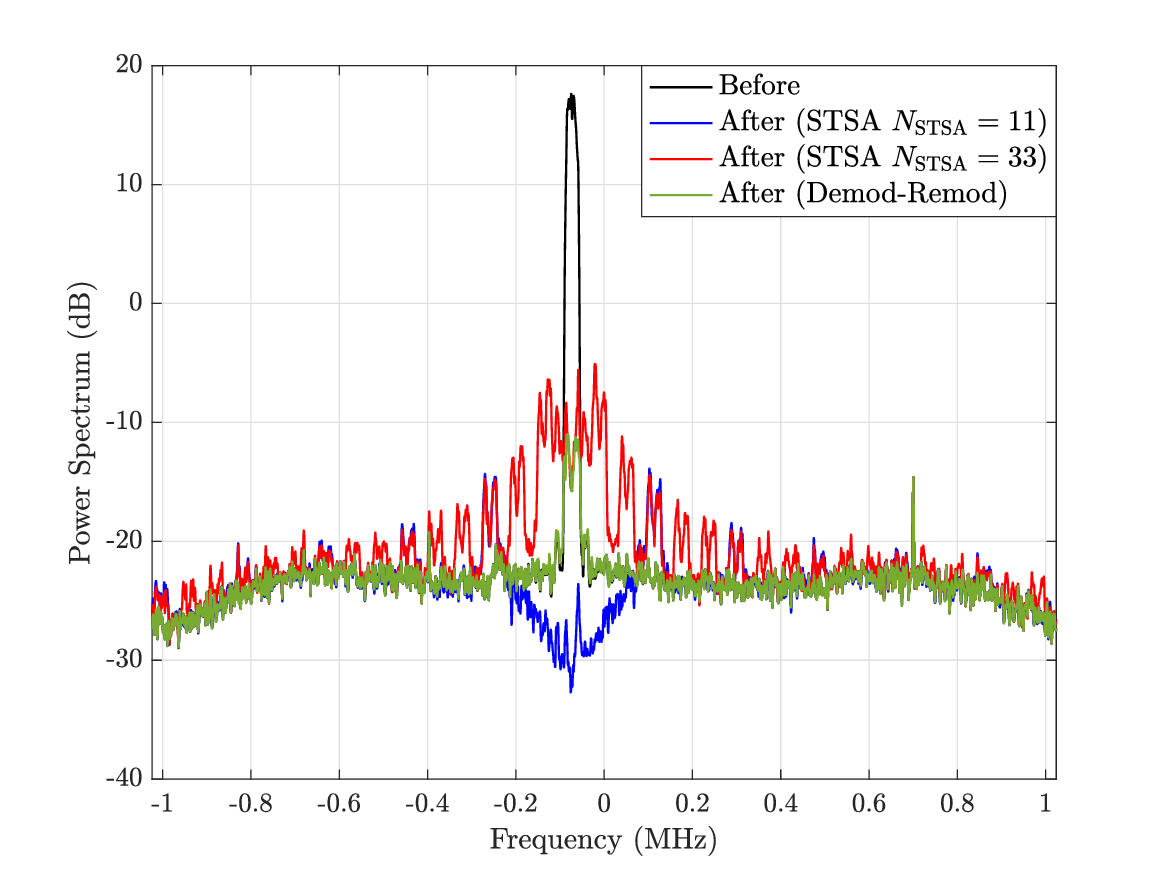}
\caption{Power spectrum for QPSK modulated signal of collected Iridium data before and after coherent subtraction with STSA and Demod-Remod approaches. \(\mathrm{INR} \approx 20.81\)dB, \(\overline{\mathrm{IRR}}_c = 20.20\)dB for STSA with \(N_{\mathrm{STSA}} = 11\), \(\overline{\mathrm{IRR}}_c = 15.20\)dB for STSA with \(N_{\mathrm{STSA}} = 33\), and \(\overline{\mathrm{IRR}}_c = 20.03\)dB for the Demod-Remod. }
\label{fig:psphigh}
\end{figure}

To further illustrate the preservation of the SOI, Fig.~\ref{fig:psplow} and \ref{fig:psphigh} present the measured power spectrum around the center frequency of the payload portion of a single Iridium burst for both the Demod-Remod and STSA methods. For moderate INRs, Fig.~\ref{fig:psplow} shows that while the STSA method effectively reduces the interference energy, it could potentially introduces spectral distortion to the SOI due to the ``scooping" effect. In contrast, the Demod-Remod approach removes less interference energy but maintains significantly better spectral integrity, preserving both in-band and out-of-band portions of the SOI. The impact of distortion becomes more pronounced at higher INR levels, as shown in Fig.~\ref{fig:psphigh}. Here, STSA creates visible degradation regardless of the block size, whereas Demod-Remod introduces minimal distortion near the interference bandwidth. Notably, it achieves comparable interference suppression to STSA with a block size of \(N_{\mathrm{STSA}} = 33\), while showing superior performance in preserving the spectrum of SOI. As no single method is optimal for all scenarios, Table~\ref{tab:1} summarizes the recommended interference mitigation techniques for each case, based on the simulation results presented above.


\begin{table*}[t!]
\centering
\caption{Selection of Interference Mitigation Techniques Under Different Scenarios. (``\cmark" indicates suitability or support for the scenario; ``\xmark" indicates lack of suitability or support for the scenario.) }
\begin{tabular}{@{}p{5.8cm}ccc@{}}
\toprule
\textbf{Scenario} & \textbf{Filtering with Ref. Antenna / MIMO} & \textbf{STSA} & \textbf{Demod-Remod} \\
\midrule
\multicolumn{4}{l}{\textbf{When Reference Antenna is Available}} \\
\midrule
High Reference INR & \cmark & \xmark & \xmark \\
Low Reference INR  & \xmark & \cmark & \cmark \\
Multiple geographically distributed antennas & \cmark & \xmark & \xmark \\
\midrule
\multicolumn{4}{l}{\textbf{When Reference Antenna is Not Available}} \\
\midrule
No Knowledge of Signal Structure       & \xmark & \cmark & \xmark \\
Partial Knowledge of Signal Structure  & \xmark & \xmark & \cmark \\
Computational Complexity               & Low    & High   & Low    \\
\bottomrule
\end{tabular}
\label{tab:1}
\end{table*}

\section{Conclusion}
\label{sec:V}
In this paper, we determined the fundamental limits of the Demod-Remod framework for removing RFI, when the interference has a narrower bandwidth or higher power than the SOI. We derived the analytical expressions for the \(\overline{\mathrm{IRR}}\) of the Demod-Remod technique in both single-carrier and OFDM systems. It was shown that while the \(\overline{\mathrm{IRR}}\) is dominated by SER and estimation errors at low INR, it is primarily limited by parameter estimation accuracy at high INR. Simulation results show that when a reference signal has low INR or not available, the Demod-Remod approach outperforms the filtering method. Moreover, assuming that the pulse shape is known, the Demod-Remod algorithm is superior to the STSA-based cancellation technique in preserving the spectrum of the SOI although the latter approach doesn't use any knowledge of the interference. 

\section*{APPENDIX}

\subsection{Derivation of \texorpdfstring{$\xi(N)$}{EN} for single-carrier Signal}
\label{app:A}
\(\xi(N)\) is defined as the MSE of the interference estimate over an \(N\) sample record. The expectation is taken over all the estimation errors of the parameters, that is \(\xi(N)=\)
\begin{equation*}
\begin{split}
&\quad \ \mathbb{E}\left[ \frac{1}{N} \sum_{n=n_0}^{n_0 + N -1} \left|Az_b(n - \epsilon)e^{j (\omega n + \theta)} - \hat{A}\hat{z}_b(n  - \hat{\epsilon}) e^{j \left(\hat{\omega} n + \hat{\theta}\right)} \right|^2 \right], \\
&= \mathbb{E}\Bigg[ \frac{1}{N} \sum_{n=n_0}^{n_0 + N -1} \left(A z_b(n - \epsilon)e^{j (\omega n + \theta)} - \hat{A} \hat{z}_b(n - \hat{\epsilon}) e^{j \left(\hat{\omega} n + \hat{\theta}\right)} \right)\\
&\qquad \ \left( A z_b(n - \epsilon)e^{j (\omega n + \theta)} - \hat{A} \hat{z}_b(n - \hat{\epsilon}) e^{j \left(\hat{\omega} n + \hat{\theta}\right)} \right)^* \Bigg], \\
&= \mathbb{E} \Bigg[ \frac{1}{N} \sum_{n=n_0}^{n_0 + N -1 } A^2|z_b(n - \epsilon)|^2 + \hat{A}^2\left| \hat{z}_b (n  - \hat{\epsilon})\right|^2 \\
&\qquad \ - A\hat{A} z_b(n - \epsilon)\hat{z}_b^*(n - \hat{\epsilon})e^{j \left((\omega - \hat{\omega}) n + \left(\theta - \hat{\theta} \right) \right)} \\
& \qquad \ -A\hat{A} z_b^*(n - \epsilon)\hat{z}_b(n - \hat{\epsilon})e^{-j \left((\omega - \hat{\omega}) n + \left(\theta - \hat{\theta} \right) \right)} \Bigg], \\
&= \mathbb{E}\Bigg[ \frac{1}{N} \sum_{n=n_0}^{n_0 + N -1 } A^2 |z_b(n)|^2 + (A + \Delta_A)^2\left| \hat{z}_b (n + \Delta_{\epsilon})\right|^2\\
&\qquad \ - 2A(A + \Delta_A)\Re\left( z_b(n)\hat{z}_b^*(n + \Delta_{\epsilon}) \right) \cos(\Delta \omega n + \Delta \theta) \Bigg],\\
\end{split}
\end{equation*}
\begin{equation}
\begin{split}
&= \mathbb{E}\Bigg[ \frac{1}{N}\sum_{n=n_0}^{n_0 + N -1 } A^2 |z_b(n)|^2 + (A^2 + \Delta_A^2)\left| \hat{z}_b (n + \Delta_{\epsilon})\right|^2 \\
& \qquad \ - 2A^2\Re\left( z_b(n)\hat{z}_b^*(n + \Delta_{\epsilon}) \right) \cos(\Delta_{\omega} n + \Delta_{\theta}) \Bigg],
\label{IRR_A1}
\end{split}
\end{equation}
where \(\Delta_A = A - \hat{A}\), and \(\Delta_{\epsilon} = \hat{\epsilon} - \epsilon\). It is easy to show that \(\frac{1}{N} \sum_n |z_b(n)|^2  = \frac{1}{N} \sum_n |\hat{z}_b(n + \Delta_{\epsilon})|^2 \approx \frac{1}{P}\mathbb{E}[|x_k|^2] \sum_n p(n)^2 \approx \frac{1}{P}\). The third term in (\ref{IRR_A1}) can be approximated using the second order Taylor expansion:
\begin{equation}
\begin{split}
&\quad \ \Re\left( z_b(n)\hat{z}_b^*(n + \Delta_{\epsilon}) \right) \\
&= \Re\left(z_b(n) \left( \hat{z}_b^*(n) + \Delta_{\epsilon} \hat{z}_b^*(n)' + \frac{\Delta_{\epsilon}^2}{2}\hat{z}_b^*(n)''\right)  \right), \\
&= \Re\left( z_b(n)\hat{z}_b^*(n) \right) + \Delta_{\epsilon} \Re \left(z_b(n)  \hat{z}_b^*(n)'\right) \\
&\quad \ + \frac{\Delta_{\epsilon}^2}{2} \Re \left( z_b(n) \hat{z}_b^*(n)'' \right).
\label{IRR_A2}
\end{split}
\end{equation}
The expectation taken over time of the first term in (\ref{IRR_A2}) becomes 
\begin{equation}
\begin{split}
&\quad \  \mathbb{E} [\Re\left( z_b(n)\hat{z}_b^*(n) \right)]  \\
&= \mathbb{E}[\Re\left( s_k p(n - kP) \hat{s}_k^* p(n - kP) \right)], \\
&= \frac{1}{P}\Re( \mathbb{E}[s_k\hat{s}_k^*]),
\end{split}
\end{equation}
where \(\mathbb{E}[s_k \hat{s}_k^*]\) denotes the correlation between the original symbol and the estimated symbol. The correlation can be written as
\begin{equation}
\begin{split}
\Re( \mathbb{E} [s_k \hat{s}_k^*]) &= \frac{\mathbb{E}\left[|s_k|^2 \right] + \mathbb{E}\left[|\hat{s}_k|^2 \right] - \mathbb{E}\left[\left|s_k - \hat{s}_k \right|^2 \right]}{2},
\label{IRR_A4}
\end{split}
\end{equation}
where
\begin{equation}
\mathbb{E}\left[\left|s_k - \hat{s}_k \right|^2 \right] = \frac{d_1^2 P_s}{2},
\end{equation}
where \(P_s\) is the SER and \(d_1\) is the average error distance between \(s_k\) and \(\hat{s}_k\), which depends on the constellation. Consequently, using the identity given in (\ref{IRR_A4}), we obtain 
\begin{equation}
\Re( \mathbb{E} [s_k \hat{s}_k^*]) = \left( 1- \frac{d_1^2 P_s}{2} \right),
\label{eq:ss_corr}
\end{equation}
and 
\begin{equation}
\mathbb{E} [\Re\left( z_b(n)\hat{z}_b^*(n) \right)] = \frac{1}{P} \left( 1- \frac{d_1^2 P_s}{2} \right).
\end{equation}
For the time expectation of the the second term in (\ref{IRR_A2}), 
\begin{equation}
\begin{split}
&\quad \ \mathbb{E}\left[\Delta_{\epsilon}  \Re \left(z_b(n)  \hat{z}_b^*(n)'\right) \right] \\
&= \Delta_{\epsilon}\mathbb{E} \left[ \Re (s_k \hat{s}_k^* p(n - kP) p'(n - kP)) \right]\,\\
&= \Delta_{\epsilon} \Re \left(  \mathbb{E}[s_k \hat{s}_k^*] \frac{1}{P} \sum_{n} p(n)p'(n) \right), \\
&= \Delta_{\epsilon} \Re \left( \mathbb{E}[s_k \hat{s}_k^*] \frac{1}{T} \int_{-\infty}^{\infty} p(t)p'(t) \ \mathrm{d}t \right), \\
&= \Delta_{\epsilon} \Re \left( \mathbb{E}[s_k \hat{s}_k^*] \left[ \frac{1}{2T} p(t)^2 \right]_{-\infty}^{\infty} \right) = 0.
\end{split}
\end{equation}
Similarly, 
\begin{equation}
\begin{split}
& \quad \ \mathbb{E}\left[ \frac{\Delta_{\epsilon}^2}{2} \Re \left( z_b(n) \hat{z}_b^*(n)'' \right) \right] \\
&= \frac{\Delta_{\epsilon}^2}{2} \Re \left( \mathbb{E}[s_k \hat{s}_k^*] \frac{1}{T} \int_{-\infty}^{\infty} p(t)p''(t) \ \mathrm{d}t \right), \\
&= -\frac{\Delta_{\epsilon}^2}{2} \Re \left( \mathbb{E}[s_k \hat{s}_k^*] \frac{1}{T} \int_{-\infty}^{\infty} p'(t)^2 \ \mathrm{d}t \right), \\
&= -\frac{\Delta_{\epsilon}^2}{2P} \left( 1- \frac{d_1^2 P_s}{2} \right)E_p',
\end{split}
\end{equation}
where \(E_p'\) denotes the energy of the differentiated pulse and is defined as \(E_p' = \sum_n |p'(n)|^2 = P^2 \sum_{n} |p(n) - p(n - 1)|^2\). Combining all, 
\begin{equation}
\mathbb{E}[\Re\left( z_b(n)\hat{z}_b^*(n + \Delta_{\epsilon}) \right) ] = \frac{1}{P} \left( 1- \frac{d_1^2 P_s}{2} \right) \left(1 - \frac{\Delta_{\epsilon}^2 E_p'}{2}  \right),
\end{equation}
If the modulation is misclassified, 
\begin{equation}
\Re\left( z_b(n)\hat{z}_b^*(n) \right) = \frac{1}{P} \left( 1- \frac{d_2^2}{2} \right) \left(1 - \frac{\Delta_{\epsilon}^2 E_p'}{2}  \right),
\end{equation}
where \(d_{2}\) is the average distance between the original symbols and their closest symbols from the other modulation schemes being classified. For the remaining cosine term in (\ref{IRR_A1}), if we let \(n_0 = -(N-1)/2\), the estimation errors between the phase and frequency are independent to each other \cite{1057115} . Moreover, following the derivation in the Appendix~A of \cite{li2025parametric}, the MSE can be rewritten as
\begin{equation}
\begin{split}
\xi(N) &= \frac{2A^2}{P} + \frac{\sigma_A^2}{P} - \Bigg( P_c \left( 1- \frac{d_1^2 P_s}{2} \right) \left(1 - \frac{\sigma_{\epsilon}^2 E_p'}{2}  \right) \\
&\quad \ + (1 - P_c) \left( 1- \frac{d_2^2}{2} \right) \left(1 - \frac{\sigma_{\epsilon}^2 E_p'}{2}  \right)  \Bigg) \\
&\quad \ \cdot \frac{2A^2}{P}  \mathbb{E} \left[ \frac{1}{N} \sum_{n = -(N-1)/2}^{(N-1)/2} \cos(\Delta_{\omega} n + \Delta_{\theta}) \right],\\
&= \frac{2A^2}{P} + \frac{\sigma_A^2}{P} - \gamma  \frac{2A^2}{P N}  e^{-\frac{\sigma_{\theta}^2}{2}} \sum_{n=-(N-1)/2}^{(N-1)/2} e^{-\frac{n^2}{2}\sigma_{\omega}^2}, \\
&\approx \frac{2A^2}{P} + \frac{\sigma_A^2}{P} - \gamma  e^{-\frac{\sigma_{\theta}^2}{2}} \frac{2\sqrt{2\pi} A^2}{ PN \sigma_\omega} \mathrm{ erf} \left(\frac{N\sigma_\omega}{2\sqrt{2}} \right),
\end{split}
\end{equation}
where \(\mathrm{erf}(\cdot)\) denotes the error function and we assume that all estimation error is Gaussian distributed with zero mean and a specific variance. Specifically, we assumed that \(\Delta_A \sim \mathcal{N}(0, \sigma_A^2)\), \(\Delta_{\theta} \sim \mathcal{N}(0, \sigma_{\theta}^2)\), \(\Delta_{\omega} \sim \mathcal{N}(0, \sigma_{\omega}^2)\), \(\Delta_{\epsilon} \sim \mathcal{N}(0, \sigma_{\epsilon}^2)\), and 
\begin{equation}
\gamma  = \left( 1 - \frac{P_c d_1^2 P_s + (1 - P_c)d_2^2}{2} \right)\left(1 - \frac{\sigma_{\epsilon}^2 E_p'}{2}  \right)
\end{equation}
consists the remaining parameters being estimated for demodulating the single-carrier signal and takes the values between 0 and 1. 

\subsection{Derivation of \texorpdfstring{$\xi(N)$}{EN} for OFDM Signal}
\label{app:B}
For OFDM signal, the only change is in the correlation term of (\ref{IRR_A2}), whose expectation can be expressed as
\begin{equation}
\begin{split}
&\quad  \ \mathbb{E} \left[ \Re\left( z_b(n)\hat{z}_b^*(n + \Delta_{\epsilon}) \right) \right]\\
&= \mathbb{E} \Bigg[\Re\Bigg( \left( \frac{1}{\sqrt{PM}}\sum_{m=0}^{M - 1}s_m e^{j 2 \pi n \frac{m}{PM}} \right) \\
& \qquad \cdot \left( \frac{1}{\sqrt{PM}}\sum_{i=0}^{M - 1}\hat{s}_i e^{j 2 \pi \left(n + \Delta_{\epsilon} \right) \frac{i}{PM}} \right)^* \Bigg) \Bigg], \\
&= \frac{1}{PM}\sum_{m=0}^{M-1} \mathbb{E} \left[\Re \left(  s_m \hat{s}_m^* e^{-j 2 \pi \Delta_{\epsilon} \frac{m}{PM}} \right)  \right], \\
&= \frac{1}{PM} \sum_{m = 0}^{M-1} \mathbb{E} \left[  \Re(s_m \hat{s}_m^*)\cos\left(- 2 \pi \Delta_{\epsilon} \frac{m}{PM} \right) \right] \\
& \quad \ - \mathbb{E} \left[  \Im(s_m \hat{s}_m^*)\sin\left( -2 \pi \Delta_{\epsilon} \frac{m}{PM} \right) \right].
\label{eq:zz_corr_OFDM}
\end{split}
\end{equation}
Similarly, by combining the result of (\ref{eq:ss_corr}) with the derivation in Appendix A of \cite{li2025parametric}, the second term in (\ref{eq:zz_corr_OFDM}) vanishes, and the remaining term simplifies to
\begin{equation}
\begin{split}
&\quad  \ \mathbb{E} \left[ \Re\left( z_b(n)\hat{z}_b^*(n + \Delta_{\epsilon}) \right) \right]\\
&= \frac{1}{P} \left( 1- \frac{d_1^2 P_s}{2} \right) \frac{1}{M} \sum_{m=1}^{M-1}\mathbb{E} \left[ \cos\left( -2 \pi \Delta_{\epsilon} \frac{m}{PM} \right) \right], \\
&=  \frac{1}{P }\left( 1- \frac{d_1^2 P_s}{2} \right) \frac{P}{2\sqrt{2 \pi } \sigma_{\epsilon}} \mathrm{erf} \left( \frac{\sqrt{2} \pi \sigma_{\epsilon}}{P} \right), \\
&\approx \frac{1}{P }\left( 1- \frac{d_1^2 P_s}{2} \right) \frac{\mathrm{erf} \left(\sqrt{2} \pi \sigma_{\epsilon} \right)}{2\sqrt{2 \pi } \sigma_{\epsilon}}, 
\end{split}
\end{equation}
where the approximation is justified by the assumption that \(\sigma_{\epsilon} \ll 1 < P\). Consequently, following the same derivation as in the single-carrier case, the expression for \(\xi(N)\) remains unchanged, while the corresponding \(\gamma\) becomes
\begin{equation}
\begin{split}
\gamma &= \left( 1 - \frac{P_c d_1^2 P_s + (1 - P_c)d_2^2}{2} \right) \frac{\mathrm{erf} \left(\sqrt{2} \pi \sigma_{\epsilon} \right)}{2\sqrt{2 \pi } \sigma_{\epsilon}}.
\end{split}
\end{equation}

\section*{ACKNOWLEDGMENT}
The authors would like to thank Dr. Steven W. Ellingson for his support and for providing the data used in this paper.

\bibliographystyle{IEEEtran}
\bibliography{mybibfile}

\begin{IEEEbiography}
[{\includegraphics[width=1in,height=1.25in,keepaspectratio]{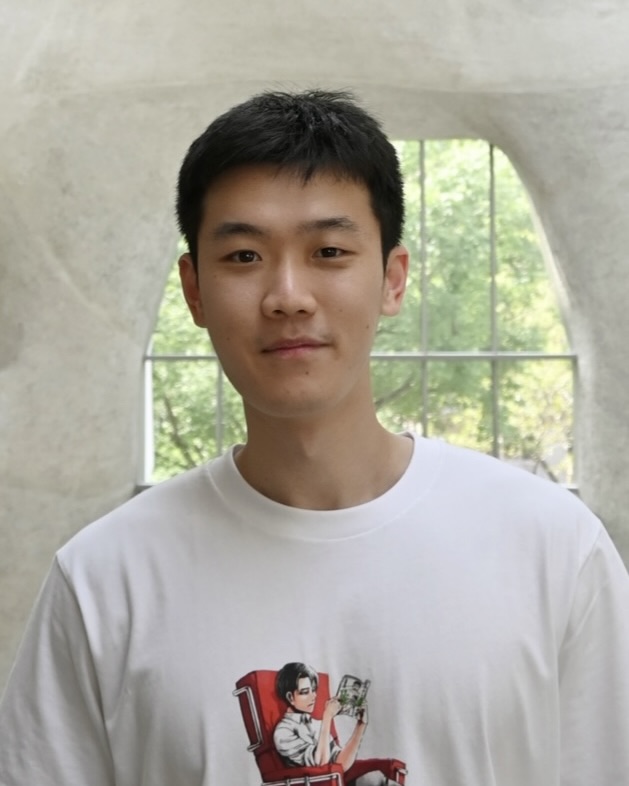}}]
{Xinrui Li} 
received the B.S. and M.S. degrees in electrical engineering from Virginia Tech, Blacksburg, VA, in 2019 and 2025, respectively, where he is pursuing the Ph.D. degree with the Bradley Department of Electrical and Computer Engineering. His research interests span the areas of communication theory and reconfigurable intelligent surface. 

\end{IEEEbiography}%

\begin{IEEEbiography}
[{\includegraphics[width=1in,height=1.25in,keepaspectratio]{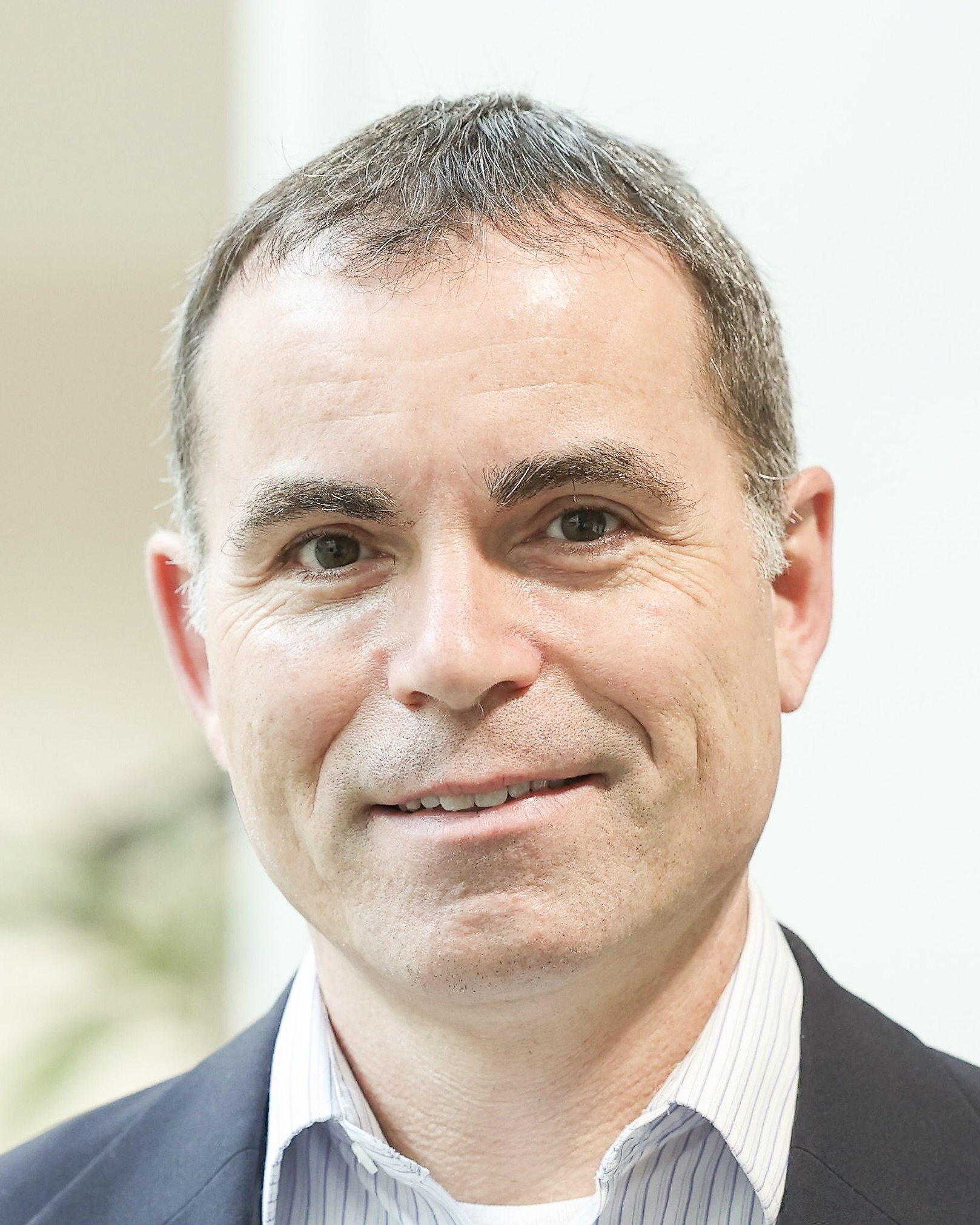}}]
{R. Michael Buehrer}{\space}(Fellow, IEEE) joined Virginia Tech from Bell Labs as an Assistant Professor with the Bradley Department of Electrical and Computer Engineering in 2001. He is currently a Professor of Electrical Engineering and is the Director of Wireless~@~Virginia Tech, a comprehensive research group focusing on wireless communications, radar and localization. During 2009, Dr. Buehrer was a visiting researcher at the Laboratory for Telecommunication Sciences (LTS) a federal research lab which focuses on telecommunication challenges for national defense. While at LTS, his research focus was in the area of cognitive radio with a particular emphasis on statistical learning techniques.

Dr. Buehrer was named an IEEE Fellow in 2016 ``for contributions to wideband signal processing in communications and geolocation.” In 2023, he received the prestigious MILCOM Lifetime Award for Technical Achievement. This award recognizes individuals who have made important technical contributions to military communications over the course of their careers. His current research interests include machine learning for wireless communications and radar, geolocation, position location networks, cognitive radio, cognitive radar, electronic warfare, dynamic spectrum sharing, communication theory, Multiple Input Multiple Output (MIMO) communications, spread spectrum, interference avoidance, and propagation modeling. His work has been funded by the National Science Foundation, the Defense Advanced Research Projects Agency, the Office of Naval Research, the Army Research Office, the Air Force Research Lab and several industrial sponsors.

Dr. Buehrer has authored or co-authored over 100 journal and approximately 300 conference papers and holds 18 patents in the area of wireless communications. In 2023, he received the prestigious MILCOM Lifetime Award for Technical Achievement, an award that recognizes individuals who have made important technical contributions to military communications over the course of their careers. In 2023 and 2021 he was the co-recipient of the Vanu Bose Award for the best paper at IEEE MILCOM. In 2023 and 2010 he was co-recipient of the Fred W. Ellersick MILCOM Award for the best paper in the unclassified technical program. He was formerly an Area Editor IEEE Wireless Communications. He was also formerly an associate editor for IEEE Transactions on Communications, IEEE Transactions on Vehicular Technologies, IEEE Transactions on Wireless Communications, IEEE Transactions on Signal Processing, IEEE Wireless Communications Letters, and IEEE Transactions on Education. He has also served as a guest editor for special issues of The Proceedings of the IEEE, and IEEE Transactions on Special Topics in Signal Processing. In 2003 he was named Outstanding New Assistant Professor by the Virginia Tech College of Engineering and in 2014 he received the Dean’s Award for Excellence in Teaching.
\end{IEEEbiography}%

\end{document}